\documentclass[aps,prb,twocolumn,superscriptaddress,bibnotes]{revtex4-1}
\usepackage{color}
\usepackage{graphicx}

\begin{document}


\title{Observation of the intrinsic magnetic susceptibility \\
of highly purified single-wall carbon nanotubes}


\author{Yusuke Nakai}
\email{nakai@tmu.ac.jp}
\affiliation{Department of Physics, Graduate School of Science and Engineering, Tokyo Metropolitan University, Tokyo 192-0397, Japan,}

\author{Ryo Tsukada}
\affiliation{Department of Physics, Graduate School of Science and Engineering, Tokyo Metropolitan University, Tokyo 192-0397, Japan,}
\author{Yasumitsu Miyata}
\affiliation{Department of Physics, Graduate School of Science and Engineering, Tokyo Metropolitan University, Tokyo 192-0397, Japan,}
\affiliation{JST, PRESTO , 4-1-8 Hon-Chou, Kawaguchi, Saitama 332-0012, Japan,}
\author{Takeshi Saito}
\author{Kenji Hata}
\affiliation{Nanotube Research Center, AIST, 1-1-1 Higashi, Tsukuba 305-8565, Japan}
\author{Yutaka Maniwa}
\email{maniwa@phyes.se.tmu.ac.jp}
\affiliation{Department of Physics, Graduate School of Science and Engineering, Tokyo Metropolitan University, Tokyo 192-0397, Japan,}

\date{\today}

\begin{abstract}
We report the observation of the intrinsic magnetic susceptibility of highly purified SWCNT samples prepared by a combination of acid treatment and density gradient ultracentrifugation (DGU). We observed that the diamagnetic susceptibility of SWCNTs increases linearly with increasing nanotube diameter. We found that the magnetic susceptibility divided by the diameter is a universal function of the scaled temperature. Furthermore, the estimated magnetic susceptibilities of pure semiconducting and pure metallic SWCNT samples suggest that they respond differently to changes in carrier density, which is consistent with theory. These findings provide experimental verification of the theoretically predicted diameter, temperature, and metallicity dependence of the magnetic susceptibility.
\end{abstract}

\pacs{75.75.-c,81.07.De}

\maketitle

The magnetism of carbon-based materials (i.e. graphene, graphite, carbon nanotubes, and fullerenes) has recently been the subject of intense research.~\cite{Yazyev_RepProgPhys2010} These materials exhibit exotic magnetism such as ferromagnetism above room temperature, which is believed to be induced by defects in the graphitic network. Our understanding of the intrinsic magnetism in these materials without such intentionally introduced defects, however, has also been very limited due to magnetic impurities remaining in samples. 
In the present work, we investigated single-wall carbon nanotubes (SWCNTs), which are rolled-up tubes of graphene sheets that exhibit unusually anisotropic electrical and magnetic properties.~\cite{saito_physical_1998,ajiki_electronic_1993,ajiki_magnetic_1993,ajiki_magnetic_1995,lu_novel_1995} In the present paper, we focus our attention on the magnetism of SWCNTs, because despite theoretical predictions of novel magnetic features, experimental studies of SWCNT magnetism have been very limited. 

The magnetism of SWCNTs is expected to be dominated by their orbital magnetic susceptibility, which is 2 orders of magnitude higher than the spin magnetic susceptibility. The magnetic susceptibility is strongly anisotropic: it shows a large diamagnetic response in a magnetic field perpendicular to the tube axis ($\chi_{\rm \perp}$). In contrast, the magnetic susceptibility in a magnetic field parallel to the tube axis ($\chi_{\rm \parallel}$) depends on whether the SWCNT is metallic or semiconducting. Approximately one-third of SWCNTs are metallic and the rest are semiconducting, depending on the chirality. The cylindrical shape of the SWCNT leads to an Aharonov-Bohm effect when a magnetic field is introduced parallel to the tube axis,~\cite{ajiki_magnetic_1993,ajiki_magnetic_1995,lu_novel_1995} resulting in a paramagnetic (diamagnetic) $\chi_{\rm \parallel}$ for metallic (semiconducting) SWCNTs. 
Actually, the predicted large magnetic susceptibility anisotropy $\Delta\chi$ =  $\chi_{\rm \parallel} - \chi_{\rm \perp}$ was estimated indirectly by magneto-optical experiments of aligned SWCNT sample in high magnetic fields.~\cite{searles_large_2010,torrens_measurement_2006,islam_magnetic_2005,zaric_estimation_2004} 
Furthermore, the magnetic susceptibility for SWCNTs depends linearly on the nanotube diameter $d$, and that there is universal scaling in scaled magnetic susceptibility $\chi/d$ as a function of scaled temperature $k_BT/\Delta_0$, where $\Delta_0$ is the characteristic energy, and corresponds to the bandgap for a semiconducting SWCNT.~\cite{lu_novel_1995} 

Magnetic susceptibility measurements of SWCNTs, however, are extremely difficult. Available SWCNT samples, which are synthesized using conventional methods such as a laser ablation, chemical vapor deposition (CVD), or arc discharge, normally contain ferromagnetic catalyst particles such as Fe, Co, and/or Ni-based compounds as well as carbonaceous impurities. These impurities are usually difficult to completely remove. Therefore, even after the usual purification processes, remaining impurities may obscure the small intrinsic magnetic susceptibility of SWCNTs, which is on the order of $10^{-6}$ emu/g. A reliable determinations of susceptibility therefore requires a method for producing relatively large quantities ($\sim$10 mg) of pure SWCNTs. 

Probably because of these difficulties, there have only been a few reports of the direct measurements of the magnetic susceptibility of SWCNTs.~\cite{kim_high-purity_2007,wu_removal_2009} In contrast, there are several reports describing the  magnetic susceptibility of multi-wall carbon nanotubes (MWCNTs) because MWCNTs can be grown without magnetic catalysts.~\cite{likodimos_magnetic_2003,lipert_nonmagnetic_2009,ramirez_magnetic_1994} For instance, Kim {\it et al}. succeeded in obtaining a diamagnetic response from a purified sample consisting of a mixture of semiconducting and metallic SWCNTs by performing air oxidation and chemical treatment with magnetic gradient filtration.~\cite{kim_high-purity_2007} Unfortunately, their results are difficult to reconcile with theory because their magnetic susceptibility of $\sim-5\times10^{-6}$ emu/g is an order of magnitude higher than the theoretical prediction.~\cite{lu_novel_1995} This discrepancy may be due to contributions from carbonaceous impurities remaining in their samples. 
Moreover, diameter, temperature and metallicity dependence of the magnetic susceptibility have yet been investigated. 
Therefore, systematic measurements of purified SWCNTs with controlled diameters and metallicity would be extremely useful in examining the theoretical calculations. 

Recently, effective post-synthesis purification techniques for bulk quantities of SWCNTs  have been developed, such as centrifugation,~\cite{nishide_effective_2009} density gradient ultracentrifugation (DGU),~\cite{arnold_sorting_2006,yanagi_optical_2008} and gel chromatography.~\cite{liu_large-scale_2011,tanaka_high-yield_2008} These techniques enable the reduction of impurity concentration, sorting by diameter and/or chirality, and separation of metallic and semiconducting SWCNTs. In the present work, a purification process combined with acid treatment and DGU was used to successfully obtain a large quantity of high-purity SWCNTs. 
The diamagnetic susceptibility of the purified sample agreed quite well with theoretical predictions. The results reinforce our general understanding of SWCNT magnetism, and will be helpful in further studies of the magnetism of SWCNTs and related nanocarbon materials. 

We used four different types of pristine SWCNTs: arc-discharge (Arc-SO, Meijo Nano-carbon Co.), high-pressure carbon monoxide (HiPco, Nano Integris), e-DIPS (enhanced direct injection pyrolytic synthesis),~\cite{saito_selective_2008} and water-assisted or super-growth chemical vapor deposition (CVD).~\cite{hata_water-assisted_2004} 
A pristine SWCNT sample was first ultra-sonicated with methanol. The solution was then vacuum-filtered to prepare a SWCNT film or buckypaper. The film was annealed at 300$^{\circ}$C for 30 minutes in air, and then immersed in 36 w/v\% HCl for 24 h at room temperature, and rinsed with deionized water. The acid immersion procedure was cycled (typically twice) until no further changes in the color of the HCl were observed. 
Then, the film was further annealed at 400$^{\circ}$C for 30 minutes in air. 
28 mg of the film was dispersed in 28 ml of 1 w/v\% deoxycholate sodium salt (DOC, Tokyo Chemical Industry Co.) solution using a bath-type ultrasonic cleaner (Sharp Co., UT-206H). The solution was dispersed using a digital sonifier (Branson, 250DA) for 8 hours at 20 \% output. The dispersed solution was centrifuged for 30 minutes at 50,000 rpm (CS100GXII, Hitachi Koki Co.). The upper 90 \% of the supernatant was collected and treated for 9 h by the DGU method with 20 w/v\% Optiprep (60\% iodixanol solution, COSMO BIO Co.). 
In this DGU process, a density gradient is formed in a centrifuge tube under strong centrifugation. 
Bundles and impurities are known to sediment lower in the gradient, and individually suspended tubes are present in the supernatant. 
The upper 20\% of the supernatant was collected and vacuum-filtered to prepare a SWCNT film. The film was annealed at 500$^{\circ}$C for 10 minutes in vacuum to remove absorbed alcohol and oxygen gas from the purified sample. 

We prepared a large quantity of SWCNT sample in order to accurately determine its magnetic susceptibility. About 10 mg of purified SWCNT film, which was covered by approximately 30 mg of aluminum foil, was sealed in a nonmagnetic quartz tube with He gas at 100 Torr. Note that the purified SWCNT film was fragmented into small pieces to achieve a random orientation of tubes. Magnetic measurements were performed in a SQUID magnetometer (MPMS, Quantum design). 

A (1 0) X-ray diffraction (XRD) peak and an S$_{22}$ (or M$_{11}$) optical absorption peak provide the average SWCNT diameter in the sample. Powder XRD was performed at the BL8B station in the Photon Factory facility, KEK, Japan. There was no trace of graphitized carbon or ferromagnetic impurities in the XRD profiles. Optical absorption spectra were measured using a UV-vis spectrophotometer (Shimadzu UV-3600). Both techniques consistently gave average diameter values $d$ (see Table 1). For the super-growth sample, we used an average diameter of $d$ = 2.65 nm, which was determined using high-resolution transmission electron microscopy (TEM).~\cite{hata_water-assisted_2004}

Figure 1 shows magnetization curves at 300 K for four different purified SWCNT samples with average diameters ranging from 1.0 to 2.65 nm. 
Residual ferromagnetic impurities provide increased magnetization in low magnetic fields, and a diamagnetic response was observed above 2 T. 
We estimated the saturation moment of the ferromagnetic impurities by subtracting a linear diamagnetic magnetization above 2 T. 
The amount of ferromagnetic impurities ($\sim10^{-2}$ emu/g) was two orders of magnitude smaller than in pristine SWCNT samples. 
In particular, a purified sample with $d$ = 1.9 nm showed a nearly diamagnetic response with a very low concentration of residual ferromagnetic impurities ($2.1\times10^{-4}$ emu/g). 
These results demonstrate that purification combined with acid treatment and DGU is extremely effective for obtaining high-purity SWCNTs. 
We also found that an as-grown super-growth sample with $d$ = 2.65 nm  after annealing at 500$^{\circ}$C for 10 min in air was already magnetically clean with a low concentration of ferromagnetic impurities ($\sim10^{-2}$ emu/g). 
This observation is consistent with TEM and thermo-gravimetric analysis (TGA) measurements.~\cite{hata_water-assisted_2004} 
\begin{figure}
\includegraphics[width = 7.5cm]{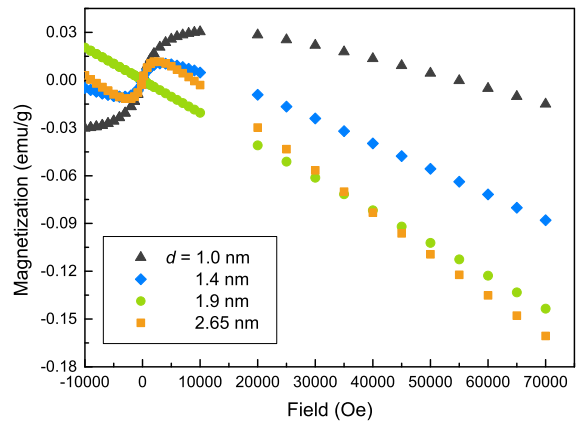}
\caption{Field dependence of the magnetization for SWCNT samples with different diameters at 300 K. Linear diamagnetic responses gave a dominant contribution at high fields while a contribution from ferromagnetic impurities was observed at low fields. The absolute value of the diamagnetic susceptibility, which is seen as a negative slope, increases with increasing nanotube diameter.\label{fig1}}
\end{figure}
\begin{table*}[tb]
\begin{center}
\begin{tabular}{|c|c|c|c|c|c|} \hline
 & $d$ (nm) & $M_s$ (emu/g) & $\chi_{\rm SWCNT}$ (emu/g) & $C$ (emu K/g) & $\theta$ (K) \\\hline\hline
HiPco & 1.0 & 5.2$\times10^{-2}$ & -1.11$\times10^{-6}$ & 4.54$\times10^{-5}$ & -14.0 \\\hline
Arc SO & 1.4 & 2.2$\times10^{-2}$ & -1.62$\times10^{-6}$ & 1.76$\times10^{-5}$ & -13.3 \\\hline
e-DIPS & 1.9 & 2.1$\times10^{-4}$ & -2.18$\times10^{-6}$ & 2.13$\times10^{-6}$ & -10.7 \\\hline
super-growth & 2.65 & 2.2$\times10^{-2}$ & -3.10$\times10^{-6}$ & 2.42$\times10^{-6}$ & 1.59 \\\hline
Arc SO (semi-rich) & 1.4 & 4.2$\times10^{-2}$ & -1.48$\times10^{-6}$ & 2.28$\times10^{-5}$ & -9.46 \\\hline
Arc SO (metal-rich) & 1.4 & 5.6$\times10^{-2}$ & -1.01$\times10^{-6}$ & 5.37$\times10^{-5}$ & -18.7 \\\hline
\end{tabular}
\end{center}
\caption{Obtained fitting parameters for the $M-T$ curves shown in Fig.2. $Ms$, $\chi_{\rm SWCNT}$, $C$, and $\theta$ represent the saturated ferromagnetic moment, intrinsic magnetic susceptibility for SWCNTs, the Curie constant, and the Weiss temperature, respectively.}
\label{}
\end{table*}

Two extrinsic magnetic susceptibilities remain in the SWCNT samples: ferromagnetic and paramagnetic impurities. These can arise from magnetic catalysts and/or carbonaceous materials. 
Defects and short SWCNTs may also result in ferromagnetism.~\cite{alexandre_edge_2008,okada_ferromagnetic_2006,okada_nanometer-scale_2003} 
Because the magnetization of ferromagnetic impurities saturates above 2 T, we can obtain magnetic susceptibility without a ferromagnetic contribution by subtracting the magnetization at 2 T from that at 6 T, and then dividing the result by 4 T. 
Figure~2 shows the obtained temperature dependence of the magnetic susceptibility without a ferromagnetic contribution. 
Paramagnetic impurities provide a Curie-Weiss contribution to the magnetization, which results in an increased magnetization at low temperatures. 
In fact, the magnetic susceptibility (closed symbols in Fig.~2) agreed quite well with the following equation: 
\begin{equation}
 \chi(T) = \chi_{\rm SWCNT} + \frac{C}{T-\theta},
\end{equation}
where $\chi_{\rm SWCNT}$ is the intrinsic magnetic susceptibility of the SWCNTs, $C$ is the Curie constant, and $\theta$ is the Weiss temperature, as summarized in Table 1. 
Because $\chi_{\rm SWCNT}$ for the $d$ = 2.65 nm sample had a strong temperature dependence particularly at high temperatures (See Fig. 2), fitting of the raw data (closed symbols) to the equation was obtained successfully below 90 K. 
Such a strong temperature dependence of $\chi_{\rm SWCNT}$ is expected for large-diameter SWCNTs (see discussion below). 
\begin{figure}
\includegraphics[width = 7.5cm]{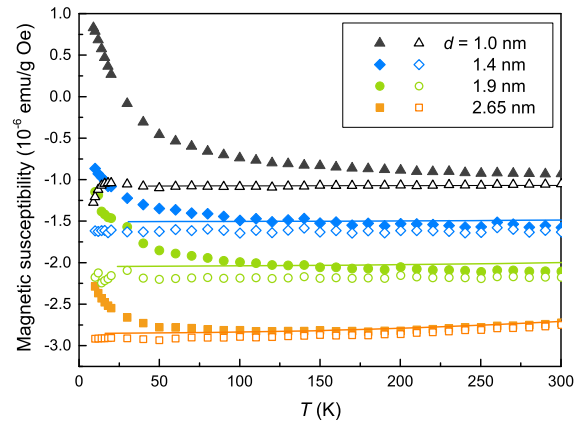}
\caption{Temperature dependence of the magnetic susceptibility. 
The solid symbols represent raw experimental data consisting of intrinsic diamagnetism and paramagnetic impurities. 
The open symbols represent the estimated diamagnetic susceptibility for the SWCNT, which was obtained by subtracting a Curie term from the raw data. The solid lines represent theoretical calculations with $|E_{\rm F}| = 0.2\Delta_0$, which agree well with the experimental data (see text). 
Note that the typical error in reading calculated data (solid line) from Ref.6 was less than 10 \%.\label{fig2}}
\end{figure}

It has been predicted that magnetic susceptibility divided by the diameter $\chi/d$ is a universal function of the scaled temperature $k_BT/\Delta_0$, where $\Delta_0$ is the characteristic energy, and corresponds to the energy gap for a semiconducting SWCNT.~\cite{lu_novel_1995} 
Therefore, SWCNTs with a larger diameter should have a larger diamagnetic susceptibility with a strong temperature sensitivity. 
In fact, Figure 2 demonstrates that the obtained magnetic susceptibility (open symbols) exhibits a stronger temperature dependence with increasing average diameter. 
Furthermore, by using the bandgap of a semiconducting SWCNT with a diameter $d$, temperature dependence of the magnetic susceptibility can be calculated.~\cite{lu_novel_1995}. The calculations of magnetic susceptibility for a variety of diameters (solid lines in Fig. 2) agree well with the experimental data over the entire temperature range.
In these calculations, it was assumed that the sample was a randomly-oriented mixture of semiconducting and metallic SWCNTs with a Fermi level $|E_{\rm F}| = 0.2 \Delta_0$, and the semiconducting to metallic SWCNT ratio was 2:1. 
Such a carrier doping is consistent with the fact that weak p-doping is likely realized in purified SWCNT samples.~\cite{nakai_giant_2014} 

As noted above, the magnetic susceptibility should have a linear dependence on nanotube diameter. 
In fact, we found a clear linear relationship between the magnetic susceptibility and the diameter in our sample, as shown in Fig. 3 (closed symbols). The observed diameter dependence of the diamagnetic susceptibility was in reasonable agreement with the calculated magnetic susceptibility (dotted line) for randomly-oriented, slightly-doped ($|E_{\rm F}| = 0.2 \Delta_0$) mixtures of SWCNTs with a 2:1 ratio of semiconducting to metallic tubes. The slight discrepancy between theory and experiment could have arisen from several sources that were neglected in both theory and experiment. On the theoretical side, for example, fine tuning of the carrier density for the different diameters used in the calculation may resolve the discrepancy. On the experimental side, although it is natural to assume that SWCNT sample is a mixture of semiconducting and metallic SWCNTs with a semiconducting to metallic tube ratio of 2:1, the actual ratio may differ slightly. Considering these factors, the reasonable agreement between experiment and theory is fairly satisfying. 
\begin{figure}
\includegraphics[width = 7.5cm]{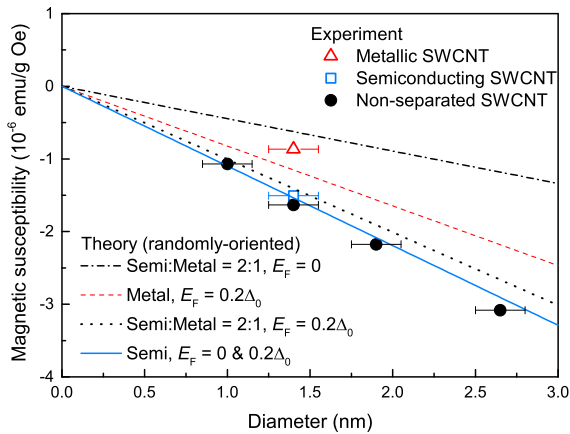}
\caption{Diameter dependence of the magnetic susceptibility. 
Note that the magnetic susceptibility at 100 K was plotted for the 2.65-nm-diameter SWCNT sample, for which the magnetic susceptibility leveled off at low temperatures. 
Solid, dashed, and dotted lines represent theoretical calculations of randomly-oriented semiconducting SWCNTs, metallic SWCNTs, and their mixtures with a ratio of semiconducting to metallic tubes of 2:1 and $|E_{\rm F}| = 0.2\Delta_0$.~\cite{lu_novel_1995} 
The dot-dash line represents mixtures with $E_{\rm F} = 0$.~\cite{lu_novel_1995}
The observed diamagnetic susceptibility of purified SWCNTs increased linearly with increasing nanotube diameter, and was consistent with theoretical calculations for mixed SWCNTs with $|E_{\rm F}| = 0.2\Delta_0$. \label{fig3}}
\end{figure}

It has been predicted that the magnetic susceptibility also depends on the electronic type of the SWCNTs:~\cite{ajiki_magnetic_1993,ajiki_magnetic_1995,lu_novel_1995} $\chi_{\parallel}$ is paramagnetic (diamagnetic), whereas $\chi_{\perp}$ is diamagnetic in undoped metallic (semiconducting) SWCNTs, resulting in paramagnetic (diamagnetic) susceptibility in randomly-oriented undoped metallic (semiconducting) SWCNT samples. 
We tested this prediction by measuring the magnetic susceptibility of semiconductor- and metal-rich SWCNT samples with $d$ = 1.4 nm prepared using a gel-chromatography method.~\cite{liu_large-scale_2011,miyata_length-sorted_2011}. 
The amount of ferromagnetic impurities was as small as in purified samples obtained by a combination of acid treatment and DGU, as shown in Table 1. Optical absorbance measurements revealed that the semiconducting to metallic ratios were 96\% and 24\% for semiconductor- and metal-rich samples, respectively. The magnetic susceptibility for pure semiconducting and pure metallic SWCNT samples, $\chi$(Semi) and $\chi$(Metal), can be estimated using the following relationship: $\chi_{\rm obs.}$ = $\alpha\chi$(Semi) + $(1- \alpha)\chi$(Metal), where $\chi_{\rm obs.}$ is the observed magnetic susceptibility for each sample and $\alpha$ is the semiconducting SWCNT fraction. 
Figure 3 shows the obtained magnetic susceptibility for pure semiconducting and metallic SWCNTs (open symbols). 
The magnetic susceptibility for randomly oriented undoped metallic SWCNTs is predicted to be positive, whereas that for semiconducting SWCNTs is predicted to be negative. 
Although $\chi$(Semi) is diamagnetic and quantitatively consistent with theoretical calculations (solid line in Fig.~3), $\chi$(Metal) is diamagnetic in contrast to theory. 
We attributed this to carrier doping effects, because even a small amount of carrier doping is expected to change $\chi_{\parallel}$ for metallic SWCNTs from paramagnetic at the charge neutral point to diamagnetic (see dashed line in Fig.~3).~\cite{ajiki_magnetic_1993,ajiki_magnetic_1995,lu_novel_1995} 
In contrast, a drastic change in $\chi_{\parallel}$ for semiconducting SWCNTs from diamagnetic to paramagnetic is expected near the first Van Hove singularities (VHSs). 
However, it is highly unlikely that the Fermi energy for purified SWCNTs would fall close to the band edges, because our previous Seebeck coefficient measurements revealed that the Fermi energy for a purified SWCNT sample was closer to the charge neutral point than the first VHSs.~\cite{nakai_giant_2014} 
Therefore, the overall experimental magnetic susceptibility for high-purity SWCNTs can be explained by the scaling nature of SWCNT magnetism that was predicted by the previous theory.~\cite{ajiki_magnetic_1993,ajiki_magnetic_1995,lu_novel_1995} 

In conclusion, we found that acid treatment combined with DGU and gel chromatography was highly effective at producing magnetically-pure SWCNT samples. The obtained samples had a diamagnetic susceptibility that increases linearly with increasing nanotube diameter. 
Furthermore, we found that the magnetic susceptibility divided by the nanotube diameter $\chi/d$ is a universal function of the scaled temperature $k_BT/\Delta_0$. 
It has been suggested that pure semiconducting and pure metallic SWCNT samples both show diamagnetism, probably due to the extreme sensitivity of metallic SWCNTs to carrier density. 
These findings are quantitatively consistent with theoretically predicted orbital magnetism and Aharonov-Bohm effects.

This work was supported in part by JSPS KAKENHI Grant Numbers 25800201 and 25246006. 

\bibliographystyle{apsrev4-1}
\bibliography{ref}

\end{document}